\documentclass[prl,showpacs,twocolumn]{revtex4}

\usepackage[dvips]{graphics}
\usepackage{amsmath, amsthm, amssymb}
\usepackage{graphicx}
\usepackage{dcolumn}
\usepackage{bm}

\newcommand{\de}[1]{\left( #1 \right)}

\newcommand{\ket}[1]{\left| #1 \right\rangle}
\newcommand{\bra}[1]{\left\langle #1 \right|}

\newcommand{\eg}{{\it{e.g.}}, }
\newcommand{\ie}{{\it{i.e.}}, }
\newcommand{\etal}{{\it{et al.}}, }

\begin{document}

\title{Useful entanglement from the Pauli principle}

\author{D. Cavalcanti$^{1}$, L. M. Moreira$^{2}$, F. M. Matinaga$^{2}$,
M. O. Terra Cunha$^{3,4}$, M. Fran\c ca Santos$^{2}$}
\email{msantos@fisica.ufmg.br}

\address{$1$ ICFO-Institut de Ciencies Fotoniques, Mediterranean
Technology Park, 08860 Castelldefels (Barcelona),Spain\\
$2$ Departamento de F\'{\i}sica - CP 702 - Universidade Federal de
Minas Gerais - 30123-970 - Belo Horizonte - MG -
Brazil\\
$3$ Departamento de Matem\'atica - CP 702 - Universidade
Federal de Minas Gerais - 30123-970 - Belo Horizonte - MG - Brazil\\
$4$ The School of Physics and Astronomy, University of Leeds,
Leeds LS2 9JT, UK}

\begin{abstract}
We address the question whether identical particle entanglement is a
useful resource for quantum information processing. We answer this
question positively by reporting a scheme to create entanglement
using semiconductor quantum wells. The Pauli exclusion
principle forces quantum correlations between the spins of two independent 
fermions in the conduction band. Selective electron-hole recombination then
transfers this entanglement to the polarization of
emitted photons, which can
subsequently be used for quantum information tasks.
\end{abstract}


\maketitle

Entanglement is both an essential key for understanding spooky
quantum phenomena~\cite{fenomenos} and a useful resource which
allows the implementation of desirable quantum information
tasks~\cite{usos}. It is then naturally important to
develop methods of extracting entanglement from physical systems as
well as using it in different applications. The usual ways of creating entanglement include
interactions~\cite{interaction}, post-selection (\eg in parametric down conversion~\cite{PDC}) or a
combination of such strategies~\cite{leb}. \

Quantum correlations naturally appear in fermionic systems. A pure state of a pair of
identical fermions is always antisymmetric, hence somehow
entangled. This entanglement comes from the indistinguishability
of the fermions and can manifest itself in one or more degrees of
freedom, depending, for example, on the spatial shape of the wave
function~\cite{Vlatko}. However, there is an interesting ongoing
debate over whether it is possible to use this strictly
spin-statistical entanglement to perform quantum information
tasks~\cite{debate}.

In this Letter we present a solution to this question, by proposing
a scheme to extract entanglement created solely by the Pauli
exclusion principle. Our scheme is quite unorthodox in the sense
that decoherence, usually viewed as an enemy for entanglement
production, actually plays an important role in the extraction
of this identical particle entanglement. 

Decoherence can usually be understood as the creation of spurious
and unavoidable correlations between the system of interest and
its environment \cite{deco}. Such correlations imply information
loss and entropy creation, with the state of the system being described 
by a density operator. When the system in question is composed, decoherence 
also tends to wash out entanglement. However, one very special
situation occurs when decoherence is dictated by a null
temperature heat reservoir, in which case, the system asymptotically approaches its lowest energy level. 
Whenever the
ground state is nondegenerate, the result of null temperature
decoherence is a pure state. For composed systems, if the
nondegenerate ground state is also entangled, decoherence can
actually create entangled states. 

Here, we describe how to combine dissipation and the Pauli Exclusion Principle to produce entangled fermions. Then, we show how to use selective recombination to extract this fermionic entanglement as useful propagating photons entangled in polarization. These general ideas apply to different fermionic systems. As an example, we describe, from now on, their application in solid state physics. In fact, recent technological development has placed semiconductors among the systems capable of producing entangled particles. For example, the decay of biexcitons in a quantum dot~\cite{XX,Benson}, or the decay of excitons in two coupled quantum dots~\cite{2QD} have been used to produce entangled photons. However, in none of these works, the Pauli Exclusion Principle is the central source of the generation of entanglement. 


Consider that a semiconductor in its ground state is excited by the promotion of two
electrons to the conduction band. These electrons will be
described by some quantum state with momentum and spin
distribution. Due to phonon scattering in a short time scale the spin state will be
essentially random (supposing no energy difference between the
possible spin polarizations). With the condition that the relaxation time
$\tau_D$ is much shorter than the recombination time $\tau_{eh}$,
the electronic momentum distribution will tend to the bottom of
the band. In fact, the quantum state will approach the lowest energy state of the 
band which, due to the Pauli Principle, corresponds to the null momentum spin
singlet. In a statistically independent way, electrons in the valence band also
relax, with the net effect of promoting the existing holes to the top of
the band. Again assuming that the relaxation time of the holes is
much shorter than the respective recombination time, there will be a singlet of
holes in the top of the valence band, as well. So far, quantum
correlations have been created by the Pauli Principle through
relaxation. The remaining question is whether these correlations
can be used to implement some quantum protocol.

\begin{figure}\centering
\begin{tabular}{cc}
  \includegraphics[scale=0.28]{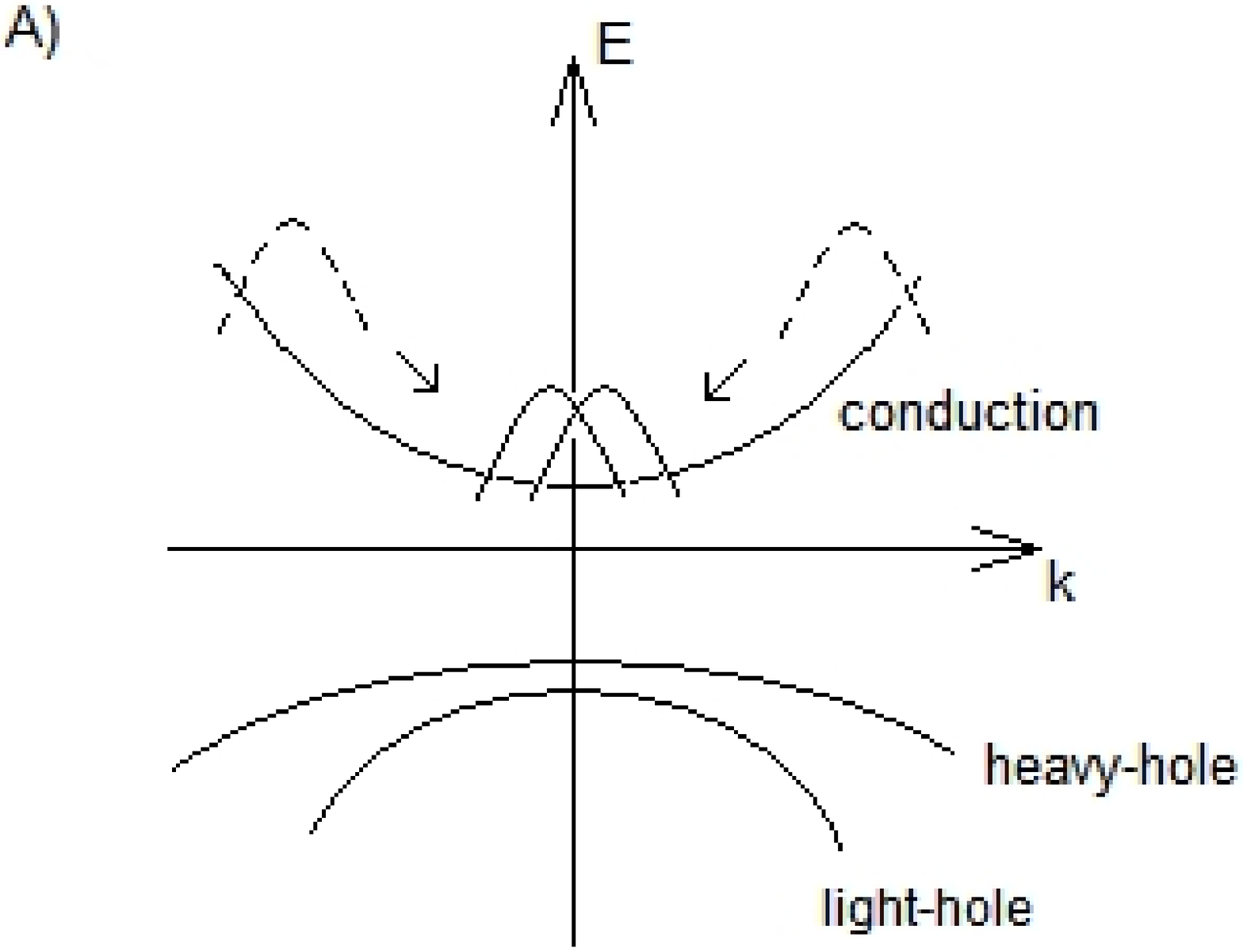}\\
  \includegraphics[scale=0.28]{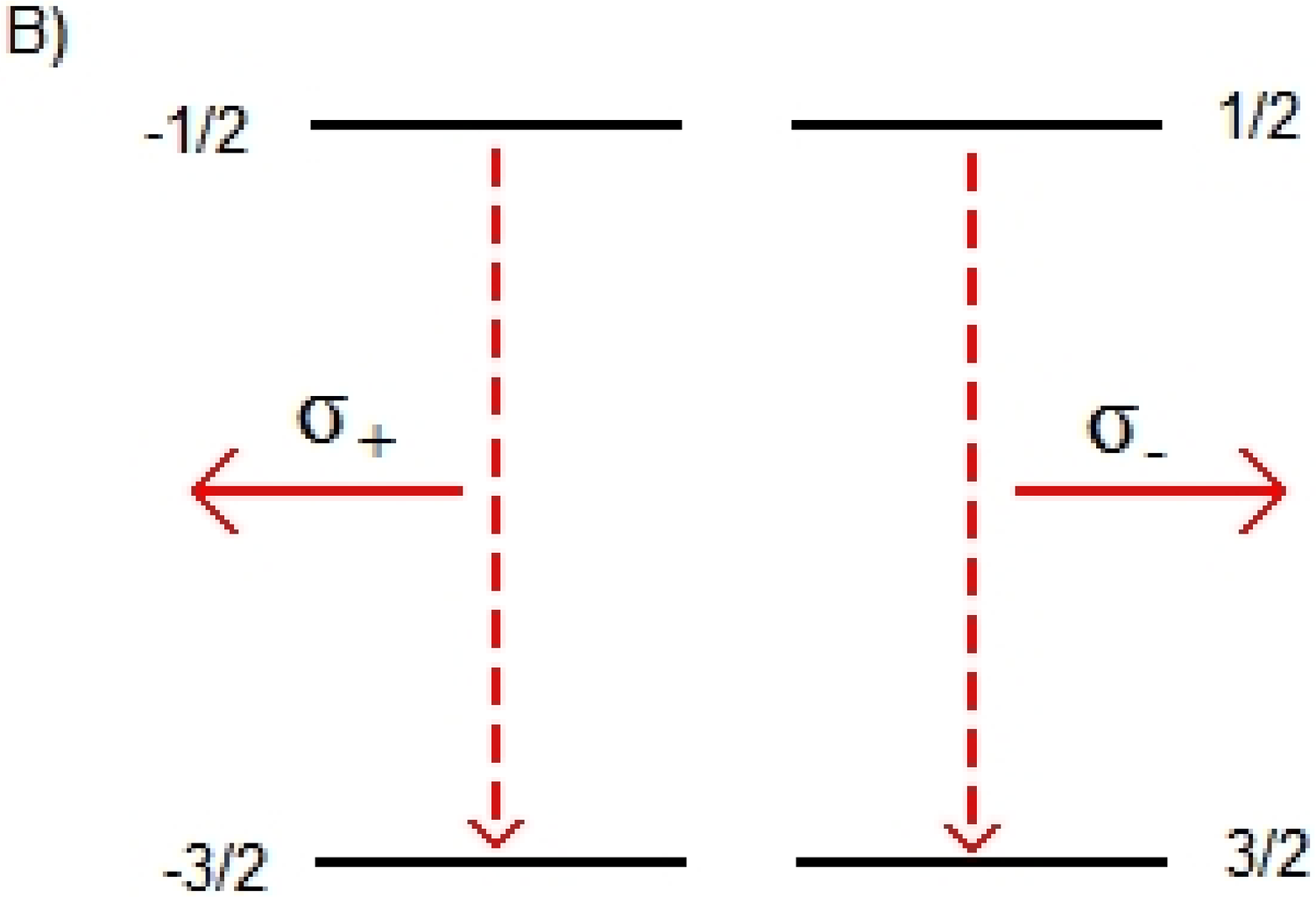}\\
  \end{tabular}
  \caption{(Color online) A) Band structure for the semiconductor quantum well. For each value of $k$ there
  can exist just two electrons according to the Pauli principle. B) Selection rules forces the transitions
  $J_{z}=-\frac{3}{2}\leftrightarrow J_{z}=-\frac{1}{2}$ through an emission/absorption of a $\sigma_{+}$ photon,
  and $J_{z}=\frac{3}{2}\leftrightarrow J_{z}=\frac{1}{2}$
  through an emission/absorption of a $\sigma_{-}$ photon.}\label{selecao}
\end{figure}

We obtain a positive answer by considering a selection rule for
the radiative electron-hole recombination: electrons with spin $+1/2$ ($-1/2$) can only
decay emitting photons circularly polarized to the right (left)
(Fig.~\ref{selecao}B). After both electrons have decayed, we
 obtain two photons in a polarization entangled state,
which can then be used for different quantum information protocols. Note that, the only effect of the
recombination is to transfer the fermionic entanglement into the emitted photons. 

Quantum Wells prepared with semiconductors of the group III-V~\cite{semic} constitute examples of physical systems that combine all the necessary ingredients. Their conduction (valence) band has
orbital angular momentum $L=0$ ($L=1$). Due to the spin-orbit interaction, the valence band 
splits into heavy-holes ($J_{z}=\pm3/2$) and light-holes ($J_{z}=\pm1/2$). The 2-D confinement lifts their degeneracy at $k=0$ producing different values for their energy gap, and, thus, allowing for the desirable selectivity (see Fig.~\ref{selecao}A). At the same time, these systems present appropriate time scales since for semiconductors, typically $\tau_D\sim 10^{-12}s$ and $\tau_{eh}\sim 10^{-9}s$.



{\bf From fermions to photons.} Let us now treat in some details
the recombination process. Suppose we have a semiconductor quantum
well with exactly two excitations with well defined momentum $k$
above the ground state (full valence band). Consider the creation
operator of two particles (electron + hole):
\begin{equation}\label{Pdag}
\Psi_{s}^{\dag}(k)=e^{\dag}_s(k) h^{\dag}_s(-k),
\end{equation}
where $e^{\dag}_s(k)$ (resp. $h^{\dag}_s(k)$) creates an electron
(hole) in the conduction (valence) band with logical spin $s$ and
momentum $k$. The spin notation is utilized to emphasize correlation
in the creation process, and the following correspondence between the
real spins with the logical basis is implied through the paper:
$$
\begin{tabular}{|c|c|c|}
  \hline
   & $\ket{0}$ & $\ket{1}$ \\
  \hline
  electrons & -1/2 & 1/2 \\
  holes & -3/2 & 3/2 \\
  \hline
\end{tabular}.
$$
The spin state we are interested in can be described by a spin density
operator for the electrons and holes determined (up to
normalization) by the correlation function \cite{Vlatko}:
\begin{eqnarray}
\rho_{{rr'ss'}}&=&\bra{\Phi_0}\Psi_{r}(k)\Psi_{r'}(k)
\Psi_{s}^{\dag}(k)\Psi_{s'}^{\dag}(k)\ket{\Phi_0}\label{rhoperf}\\
&=&{\bra{\phi^{\de{e}}_0}e_r(k) e_{r'}(k)
e^{\dag}_{s}(k)e^{\dag}_{s'}(k)\ket{\phi^{\de{e}}_0}}\nonumber\\
&\times&{\bra{\phi^{\de{h}}_0}
h_r(-k) h_{r'}(-k)h_{s}^{\dag}(-k)h_{s'}^{\dag}(-k)\ket{\phi^{\de{h}}_0}},\nonumber
\end{eqnarray}
where $\ket{\phi^{\de{e}}_0}$ $\de{\ket{\phi^{\de{h}}_0}}$ denotes
the electron (hole) initial state (vacuum) and $\ket{\Phi_0}$ is their tensor
product. As the operators obey fermionic anti-commutation rules
\begin{equation}\label{anticom}
[e_s^{\dag}(k),e_{s'}(k')]_{+}= \delta_{ss'}\delta(k-k')
\end{equation}
(the same for  $h_s(k)$), we have
\begin{equation}
\rho_{{rr'ss'}}=\de{\delta_{rs'}\delta_{r's}-\delta_{rs}\delta_{r's'}}^2.
\end{equation}
Note that we use a shortened label to represent the matrix
elements \eqref{rhoperf}. In the electron-hole-electron-hole
ordering, this operator is a density matrix representing the
(unnormalized) state
\begin{equation}\label{perfect state}
\ket{\psi}=\ket{-\frac{1}{2},-\frac{3}{2};+\frac{1}{2},+\frac{3}{2}}+\ket{+\frac{1}{2},+\frac{3}{2};-\frac{1}{2},-\frac{3}{2}}.
\end{equation}

In order to enhance the emission process, and also to have control
over the emitted photons, the sample can be placed within an optical
cavity~\cite{cavidade} in resonance with the transition between the top the heavy-hole valence band and the bottom of the conduction band. Following these selection rules, the process described above will generate photon pairs in the polarization state
\begin{equation}\label{perfect photon}
{\ket{\sigma_{-}\sigma_{+}}+\ket{\sigma_{+}\sigma_{-}}},
\end{equation}
which is a maximally entangled Bell state. Note that it is essential that the electrons have the same momentum for the creation of a maximally entangled pair of photons.


\textbf{Some imperfections.} The scenario that was discussed up to
now is pretty much idealized. It is important to stress, however,
that one very robust point in favor of this proposal is its
independency on the specific model of decoherence used. Whenever
the conditions on the time scales here imposed are fulfilled, the
state of the system before recombination will be very close to the
one here described. And so will the state of the emitted photons.

One nice way to mimic the effects of imperfections in our approach is to consider
a broadening in the momentum distribution, and also an imperfect coincidence of the momenta.
This can be phenomenologically taken into account by modifying Eq.~\eqref{Pdag} to
\begin{equation}\label{criacao}
\Psi_{ss}^{\dag}(k)=\int\!\!\!\int
f(k_j-k)f(\widetilde{k}_j-\widetilde{k})e_{s}^{\dag}(k_j)h_{s}^{\dag}(\widetilde{k}_j)dk_jd\widetilde{k}_j.
\end{equation}
This operator creates an electron with spin $s$ and momentum
distribution given by the function $f(k_j-k)$, and a hole with spin
$s$ and momentum distribution given by
$f(\widetilde{k}_j-\widetilde{k})$. Later we will associate
$k=-\widetilde{k}$. With this operator, Eq.~\eqref{rhoperf} can be rewritten as
\begin{widetext}
\begin{eqnarray}\label{rho}
\rho_{{rr'ss'}}&=&\int dk_0 ... dk_3
f^*(k_0-k)f^*(k_1-k')f(k_2-k)f(k_3-k')
\bra{\phi_0^{\de{e}}}e_{r}(k_0)e_{r'}(k_1)e_{s}^{\dag}(k_2)e_{s'}^{\dag}(k_3)
\ket{\phi_0^{\de{e}}}\nonumber\\ &\times&\int d\widetilde{k}_0 ...
d\widetilde{k}_3
f^*(\widetilde{k}_0-\widetilde{k})f^*(\widetilde{k}_1-\widetilde{k}')f(\widetilde{k}_2-\widetilde{k})f(\widetilde{k}_3-\widetilde{k}')\bra{\phi_0^{\de{h}}}h_{r}(\widetilde{k}_0)h_{r'}(\widetilde{k}_1)
h^{\dag}_{s}(\widetilde{k}_2)h^{\dag}_{s'}(\widetilde{k}_3)\ket{\phi_0^{\de{h}}}.\nonumber\\
\end{eqnarray}
\end{widetext}
Anti-commutation rules \eqref{anticom} imply that the only
non-null matrix elements are:
\begin{subequations}
\begin{eqnarray}
\rho_{0000}&=&\rho_{1111}\\&=&(L(k,k')-M(k,k'))(\widetilde{L}(\widetilde{k},\widetilde{k}')-\widetilde{M}(\widetilde{k},\widetilde{k}')),\nonumber
\end{eqnarray}
\begin{equation}
\rho_{0101}=\rho_{1010}=L(k,k')\widetilde{L}(\widetilde{k},\widetilde{k}'),
\end{equation}
\begin{equation}
\rho_{0110}=\rho_{1001}=M(k,k')\widetilde{M}(\widetilde{k},\widetilde{k}'),
\end{equation}
\end{subequations}
where
\begin{subequations}
\begin{equation}
L(k,k')=\int dk_0dk_1|f(k_0-k)|^2|f(k_1-k')|^2,
\end{equation}
\begin{equation}
M(k,k')=\int dk_0 dk_1f^*(k_0-k)f^*(k_1-k')f(k_1-k)f(k_0-k'),
\end{equation}
\end{subequations}
with similar expressions for
$\widetilde{L}(\widetilde{k},\widetilde{k}')$ and
$\widetilde{M}(\widetilde{k},\widetilde{k}')$ by changing $k\mapsto
\widetilde{k}$ and $k' \mapsto \widetilde{k}'$.
\begin{figure}\centering
   \rotatebox{270}{\includegraphics[scale=0.40]{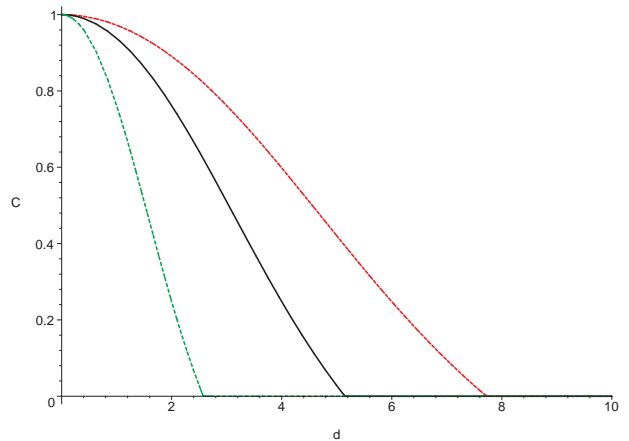}}\\
  \caption{(Color online) Photonic entanglement versus  $d=|k-k'|$.
  Green-dotted line: $\delta=2$, black-solid line: $\delta=4$,
  red-dashed curve: $\delta=6$, where $\delta$ is a measure of the
  width of the momentum distribution, assumed the same for electrons and holes (see Eq.~\eqref{distribuicao}). We see that
  the greater the spread in the momentum, the higher the
  entanglement between the photons. This is due to the fact that wider momentum
  distributions blur the difference in k's, so that once again it is impossible to
  distinguish the electrons by momentum.}\label{2X}
\end{figure}
The emitted photons
(non-normalized) polarization state  will thus be:
\begin{equation}\label{photons}
\rho=\left(%
\begin{array}{cccc}
   (L-M)(\widetilde{L}-\widetilde{M})&  &  &  \\
   & L\widetilde{L} & M\widetilde{M} &  \\
   & M\widetilde{M} & L\widetilde{L} &  \\
   &  &  & (L-M)(\widetilde{L}-\widetilde{M}) \\
\end{array}%
\right),
\end{equation}
where we leave blank the null entries. Note that when $k=k'$ and
$\widetilde{k}=\widetilde{k}'$ hold,  $L=M$ and
$\widetilde{L}=\widetilde{M}$, giving a maximally entangled
photonic state. Notice the generality of this result: state
\eqref{photons} is written in terms of arbitrary momentum
distributions of the electrons before the decaying process. For
illustration, a chart of the entanglement between the two photons
(characterized by the concurrence \cite{Conc}) versus the
difference in momentum distribution of the decaying electrons
 is displayed in Fig.~\ref{2X}. We have
chosen a Lorentzian distribution of spread $\delta$ centered in $k$
($k'$) for the momentum distributions, \ie
\begin{equation}
f(k_j-k)=\frac{\delta}{\pi[(k_j-k)^2+\delta^2]}.\label{distribuicao}
\end{equation}

\textbf{Conclusion.} From the fundamental point of view, we stress
once again that the origin of the entanglement lies in the fermionic
nature of the electrons. We claim this gives a decisive positive
answer to the question whether identical-particle entanglement is
useful for quantum information purposes. Specifically,
identical-particle entanglement can in fact be extracted and
converted into usual entanglement, and one important ingredient in
this convertibility is the use of more than one degree of freedom.
Finally, we would like to emphasize again the role played by the
coupling to the environment. Usually, decoherence is seen as the
road from quantum to classical, which makes it a plague for quantum
information tasks. Here it plays the crucial role of washing out the
distinguishable origin of the input photons, allowing the extraction
of entanglement from the Pauli principle.
%

\begin{acknowledgments}
The authors thank P.S.S. Guimar\~aes, D. Jonathan, and E. Ginossar
for useful conversations. This research was supported by CNPq,
PRPq-UFMG, and is part of the Millennium Institute for Quantum
Information project.
\end{acknowledgments}


\begin{thebibliography}{20}
\bibitem{fenomenos} A. Osterloh, L. Amico, G. Falci, and R. Fazio, Nature
{\bf{416}}, 608 (2002); S. Gosh, T. F. Rosenbaum, G. Aeppli, and
S. N. Coppersmith, \nat {\bf 425}, 48 (2003); V. Vedral, New J.
Phys. {\bf 6}, 102 (2004); N. Lambert, C. Emary, and T. Brandes,
\prl {\bf 92}, 073602 (2004).

\bibitem{usos}M.A. Nielsen and I.L. Chuang, {\emph{Quantum Computation and
Quantum Information}} (Cambridge Univesity Press, Cambridge, 2000).

\bibitem{interaction}D. P. DiVincenzo, D. Bacon, J. Kempe,  G.
Burkard, and K. B. Whaley, \nat {\bf408}, 339 (2000); W. D.
Oliver, F. Yamaguchi, and Y. Yamamoto, \prl {\bf 88}, 037901
(2002).

\bibitem{PDC} P. Kwiat \etal \prl {\bf 75}, 4337 (1995).

\bibitem{leb}A. V. Lebedev \etal \prb {\bf 69}, 235312 (2004).


\bibitem{Vlatko}V. Vedral, Cent. Eur. J. Phys. {\bf 2}, 289 (2003).
D. Cavalcanti \etal \pra {\bf{72}}, 062307 (2005); M. R. Dowling, A. C. Dohery, and H. M. Wiseman, Phys. Rev. A {\bf 73}, 052323 (2006).


\bibitem{debate}K. Eckert, J. Schliemann, D. Bruss, and M. Lewenstein, Annals of Physics {\bf299}, 88
(2002); G. Ghirardi and L. Marinatto, \pra {\bf70}, 012109 (2004).

\bibitem{deco}Giulini \etal {\it{Decoherence and the appearance of a classical world in quantum theory}} (Springer, Berlin,
1996).

\bibitem{Benson} O. Benson, C. Santori, M. Pelton, and Y. Yamamoto, Phys. Rev.
Lett. {\bf 84}, 2513 (2000).

\bibitem{XX}N. Akopian \etal \prl {\bf 96}, 130501  (2006);
R. J. Young \etal New J. Phys. {\bf 8}, 29 (2006);
T. M. Stace, G. J. Milburn, and C. H. W. Barnes, \prb {\bf
67}, 085317 (2003).

\bibitem{2QD}C. Emary, B. Trauzettel, and C.W. J. Beenakker, \prl
{\bf 95}, 127401 (2005); M. Titov, B. Trauzettel, B. Michaelis, and
C. W. J. Beenakker, New J. Phys. {\bf 7}, 186 (2005); V. Cerletti,
O. Gywat, and D. Loss, \prb {\bf 72}, 115316 (2005).

\bibitem{semic}P. Y. Yu and M. Cardona, {\it Fundamentals of Semiconductors} (Springer-Verlag, Berlin, 1996).

\bibitem{cavidade}Y. Yamamoto, {\it{Coherence, amplification, and quantum effects in semiconductor lasers}} (John Wiley Sons, New York, 1991);
C. Weisbuch, M. Nishioka, A. Ishikawa, and Y. Arakawa, \prl {\bf
69}, 3314 (1992);G. Bj\"{o}rk, S. Machida, Y. Yamamoto, and K.
Igeta, \pra {\bf 44}, 669 (1991).





\bibitem{Conc}W. K. Wootters, \prl {\bf80}, {2245} (1998).


\end{thebibliography}
\end{document}